\documentclass[conference]{IEEEtran}
\IEEEoverridecommandlockouts

\usepackage{cite}
\usepackage{overpic}
\usepackage{comment}
\usepackage{amsmath,amssymb,amsfonts}
\usepackage{graphicx}
\usepackage{textcomp}
\usepackage{xcolor}
\usepackage{float}
\usepackage{amsthm}
\usepackage{url}
\usepackage{graphicx}
\usepackage{epstopdf}
\usepackage{amsmath,bm}
\usepackage{amsfonts}
\usepackage{amssymb}
\usepackage{color}
\usepackage{multirow}
\usepackage{multicol}
\usepackage{soul,xcolor}
\usepackage{algorithm}
\usepackage{algpseudocode}%

\usepackage{comment}
\usepackage{booktabs}
\usepackage{multirow}  
\usepackage{subcaption}

\theoremstyle{plain}

\newcommand{\vect}[1]{\mathbf{#1}}

\def\diag{\mathrm{diag}}

\def\tr{\mathrm{tr}}

\def\Htran{\mbox{\tiny $\mathrm{H}$}}
\def\Ttran{\mbox{\tiny $\mathrm{T}$}}
\def\CN{\mathcal{N}_{\mathbb{C}}} 
\def\imagunit{\mathsf{j}} 

\begin{document}

\title{Learning Energy-Efficient Modular Arrays under Hardware Non-linearities 
\thanks{The work by \"O. T. Demir was supported by 2232-B International Fellowship for Early Stage Researchers Programme funded by the Scientific and Technological Research Council of T\"urkiye. }}

\author{\IEEEauthorblockN{ \"Ozlem Tu\u{g}fe Demir$^*$ and Alva Kosasih$^{\dagger}$} 
\IEEEauthorblockA{{$^*$Department of Electrical and Electronics Engineering, Bilkent University, Ankara, Turkiye
		}\\ {$^\dagger$Nokia Standards, Espoo, Finland 
		}  \\
		{Email: ozlemtugfedemir@bilkent.edu.tr, alva.kosasih@nokia.com   }
}
}

\maketitle

\begin{abstract}
This paper investigates the joint optimization of power allocation and antenna activation in sparse extremely large aperture array systems operating under power amplifier non-linearities. We first derive an analytical expression for the achievable spectral efficiency (SE) of point-to-point MIMO channels affected by non-linear distortions using the Bussgang decomposition. To address the combinatorial and non-convex nature of the energy-efficiency (EE) maximization problem, we employ an unsupervised deep neural network (DNN) that learns the non-linear mapping between the channel state information and the optimal EE operating point. The DNN jointly predicts distortion-aware power allocation, total transmit power scaling, and modular sub-array activation based on singular-value and geometric channel features. Numerical results demonstrate that the proposed DNN-based arrays achieve significant EE gains over the conventional sparse arrays.
\end{abstract}

\begin{IEEEkeywords} ELAA, modular array, hardware impairments, MIMO capacity
\end{IEEEkeywords}

\vspace{-2mm}
\section{Introduction}
Deploying a large number of antennas over a large surface places the transmitter and receiver in the near-field region, as expected in 6G and beyond. This gives rise to the beamfocusing, where the beam concentrates on specific locations. The beamfocusing enables the base station (BS) to serve more user equipments (UEs) by allowing multiplexing not only in the angular domain but also in the distance domain. This new multiplexing paradigm significantly enhances spectral efficiency (SE), as more MIMO layers can be supported \cite{9903389,2023_Ramezani_Bits}. However, these gains come at the cost of high complexity and power consumption, driven by the need to process a massive number of antennas. 

A natural way to maintain a large aperture while reducing processing complexity is to use a sparse array \cite{10734395,2024_Zhou_Arxiv,Wang2023_GCWKS,2024_Chen_JSTSP,kosasih2025near}. This array can be viewed as a standard half-wavelength array with some deactivated antennas, leaving active elements spaced intermittently. Among various sparse configurations, modular arrays have been shown in \cite{kosasih2025near} to achieve a desirable beamfocusing pattern with a less number of antennas.

Recent discussions within the Third Generation Partnership Project (3GPP) on next-generation MIMO for 6G emphasize reducing power consumption and signaling overhead by muting or deactivating antenna ports. This direction is supported by technical documents from the 3GPP 6G Workshop held in March 2025 \cite{Nokia_6GWS,Ericsson_6GWS}. These reports further highlight a shift in key performance indicators—from focusing solely on SE, (i.e., high throughput), to prioritize energy efficiency (EE).

We draw inspiration from the use of sparse arrays and antenna port deactivation to preserve the SE advantages of near-field multiplexing while improving EE. In particular, while sparse arrays already maintain the same aperture with fewer antennas, we further enhance efficiency by selectively deactivating additional antennas within the sparse array. However, several challenges arise in realizing such systems. First, a computationally intensive optimization to determine which antennas to deactivate, resulting in high processing complexity and limited practicality. Second, hardware impairments, particularly power amplifier non-linearities \cite{9072380, Bjornson2019e}, further complicate SE optimizations.

In this paper, we propose an energy-efficient near-field MIMO solution based on modular arrays, formed through selectively deactivating antennas within a planar array. The major contributions of the paper are summarized as follows:
\begin{itemize}
    \item We develop a novel optimization framework that determines the optimal structure of a modular array. The modular array consists of four identical sub-arrays located at the corners of the structure. The goal is to determine the number of active antennas along the horizontal and vertical directions in each sub-array using a deep neural network (DNN) trained in an unsupervised manner based on the propagation environment. For example, for a $6\times6$ array, the DNN outputs activation parameters $(2,2)$, indicating that only two antennas are active along each direction. As a result, the array is formed by four $2\times2$ sub-arrays located at the corners of the modular structure. This optimization is relatively simple due to the limited set of possible outputs, unlike movable or fluid antenna systems \cite{Wong2021,Zhu2025a} and conventional antenna selection schemes \cite{Gao2013,2024_Rajapaksha}, which require complex optimization for each channel realization.
    \item We derive a novel SE expression that explicitly accounts for power amplifier non-linearities in a point-to-point MIMO setup, enabling distortion-aware performance modeling. This contrasts with existing array optimization frameworks, which typically ignore such realistic hardware impairments \cite{Gao2013,2024_Rajapaksha}. 
\end{itemize}
The DNN jointly optimizes the modular array by selecting the active antennas in each sub-array and the power allocation strategy, balancing EE and SE under practical hardware constraints.

\vspace{-2mm}
\section{System Model}

We consider an extremely
large aperture array (ELAA) architecture comprising a total of $\overline{K}$ antennas arranged in a uniform planar array (UPA) configuration. The ELAA communicates with a UE equipped with $M$ antennas in the downlink. The UE is assumed to be located in the radiative near-field region of the ELAA, and a line-of-sight (LOS) propagation environment is considered. Owing to the near-field conditions, the LOS channel matrix is not necessarily of rank one, which enables spatial multiplexing. During the channel estimation phase, the UE location—and consequently the LOS channel—is estimated and assumed to be perfectly known at the ELAA.

In this paper, we focus on the hardware distortions introduced by the power amplifiers connected to each antenna element of the ELAA. This focus is motivated by the fact that a large number of low-cost antennas are deployed along the ELAA, making hardware distortions at the ELAA side more critical compared to those at the UE. Therefore, we assume ideal hardware at the UE to isolate and analyze the impact of power amplifier non-linearities at the ELAA. We consider fully digital precoding; however, due to hardware impairments and the cost associated with each RF chain, not all antennas are activated simultaneously. Instead, based on the UE location information, we develop an unsupervised learning framework to design a sparse modular array structure by selecting a subset of active antennas so as to maximize the EE.

In the following, we derive the achievable SE for a point-to-point MIMO channel under power amplifier non-linearities. 
The derived expression assumes that $K$ out of $\overline{K}$ antennas are selected based on the given UE location (i.e., channel). 
It is worth highlighting that the result is applicable to any MIMO channel affected by power amplifier non-linearities, independent of the specific array geometry.

\section{Achievable SE of MIMO Channels under Power Amplifier Non-linearities}

We let $\vect{x} \in \mathbb{C}^K$ denote the signal vector transmitted from the $K$ antennas of the ELAA. Gaussian signaling is assumed, inspired by the classical capacity-achieving scheme for MIMO systems with ideal hardware. Accordingly, we have $\vect{x} \sim \CN(\vect{0}, \vect{Q})$, where $\vect{Q}$ denotes the transmit covariance matrix. The $k$th entry of $\vect{x}$ is denoted by $x_k$. 

The power amplifier non-linearities are modeled using the standard third-order polynomial distortion model \cite{9072380}. In addition to the desired signal $x_k$ at the $k$th RF chain, there is an additional third-order distortion component scaled by the compression parameter $\rho \leq 0$. The distorted output signal of the $k$th power amplifier is thus given by
\begin{align}
    z_k &= x_k + \frac{\rho}{\mathbb{E}\{|x_k|^2\}} |x_k|^2 x_k, \nonumber\\
    &= x_k + \frac{\rho}{q_{k,k}}|x_k|^2x_k, \quad k = 1, \ldots, K,
\end{align}
where $\mathbb{E}\{|x_k|^2\}=q_{k,k}$, i.e., the $(k,k)$th entry of $\vect{Q}$. The vectorized form by collecting $z_k$'s is obtained as $\vect{z}=[z_1 \ldots z_K]^{\Ttran}$. The $\vect{z}$ is transmitted and passes through the ELAA-UE channel $\vect{H}\in \mathbb{C}^{M\times K}$ and the received signal at the $M$ antennas of the UE is given as
\begin{align}
    \vect{y}=\vect{H}\vect{z}+\vect{n}
\end{align}
where $\vect{n} \in \CN(\vect{0}, \sigma^2\vect{I}_M)$ is the additive white Gaussian noise.

To obtain a tractable term, we can use the Bussgang decomposition \cite{Demir2020a} to express $\vect{z}$ as a linear decomposition of $\vect{x}$ and an uncorrelated distortion term as
\begin{align}
    \vect{z} = \vect{B}\vect{x}+\boldsymbol{\eta}
\end{align}
where $\vect{B}$ is the Bussgang gain matrix which can be computed from \cite{Bjornson2019e} as 
\begin{align}
    \vect{B} = (1+2\rho)\vect{I}_K.
\end{align}

The correlation matrix of the distortion $\boldsymbol{\eta}$ is computed from \cite{Bjornson2019e} as
\begin{align}
\vect{C}_{\eta}&=2\rho^2\overline{\vect{Q}}^{-1}\left(\vect{Q}\odot \vect{Q}^{*}\odot \vect{Q}\right)\overline{\vect{Q}}^{-1}.
\end{align}
where $\overline{\vect{Q}}=\diag(q_{1,1},q_{2,2},\ldots,q_{K,K})$. The transmitted power can be computed as $\tr\left(\vect{C}_z\right)$, where $\vect{C}_z=\mathbb{E}\{\vect{z}\vect{z}^{\Htran}\}$. From \cite{Bjornson2019e}, we have
\begin{align}
\tr\left(\vect{C}_z\right)=(1+2\rho)^2\tr(\vect{Q})+\tr(\vect{C}_{\eta}).
\end{align}

Using Bussgang decomposition, we can write the received signal as
\begin{align}
    \vect{y}=(1+2\rho)\vect{H}\vect{x} + \vect{H}\boldsymbol{\eta}+\vect{n},
\end{align}
where the effective noise $\overline{\vect{n}}\triangleq \vect{H}\boldsymbol{\eta}+\vect{n}$ is uncorrelated with $\vect{x}$ and it has the covariance matrix $\vect{C}_{\overline{n}}=\vect{H}\vect{C}_{\eta}\vect{H}^{\Htran}+\sigma^2\vect{I}_M$. Using then the worst-case uncorrelated additive noise theorem \cite{hassibi2003much}, we can compute a lower bound to the capacity for UE, given as
\begin{align}
    \mathrm{SE} = \log_2\left(\det\left(\vect{I}_M+(1+2\rho)^2\vect{C}_{\overline{n}}^{-1}\vect{H}\vect{Q}\vect{H}^{\Htran}\right)\right). \label{eq:SE}
\end{align}

\section{Power Consumption and EE}
In this section, we quantify the total radio power consumption of the ELAA and evaluate its corresponding EE, defined as the number of transmitted bits per unit of energy (in bits/Joule). EE is a key performance metric for 6G communications, particularly in the radiative near-field region, where UEs already experience high data rates. Instead of pursuing the highest possible SE, it is more sustainable to operate at the most energy-efficient point that balances SE and power consumption. This perspective is especially relevant from an operator’s standpoint, as it promotes greener and more cost-effective network operation.

Following the model in \cite{enqvist2024fundamentals}, the total radio power consumption of the BS array can be expressed as
\begin{align}
P_{\rm tot} = \frac{\tr(\vect{Q})}{\kappa} + \mu + (D_0 + \upsilon B)K + \eta B \cdot \mathrm{SE}, \label{eq:Ptot}
\end{align}
where $\tr(\vect{Q})$ denotes the total input power to the power amplifiers, and $0 < \kappa \leq 1$ represents the power amplifier efficiency. When $\kappa$ is strictly less than one, as in practice, the amplifiers consume more power than their input signal power.

The term $\mu$ accounts for the fixed, load-independent power consumption required for components such as cooling systems, control signaling, backhaul infrastructure, baseband processors, and local oscillators. The parameter $D_0$ represents the power consumed by each RF chain associated with a single antenna port, which includes the power used by converters, mixers, and filters.

The coefficient $\upsilon$ captures the processing energy consumption per sample due to operations such as channel estimation and precoding. When multiplied by the bandwidth $B$ (equivalently, the symbol rate), it gives the processing power consumption per antenna. Thus, with $K$ active antennas, the total RF chain and processing power consumption is $(D_0 + \upsilon B)K$.

Finally, $\eta$ characterizes the energy consumption related to data coding operations, and its contribution scales with the achievable data rate $B \cdot \mathrm{SE}$.

Given the total power consumption $P_{\rm tot}$, the main performance metric—EE—is defined as
\begin{align}
\mathrm{EE} = \frac{B \cdot \mathrm{SE}}{P_{\rm tot}},
\end{align}
where the achievable SE from \eqref{eq:SE} is used. The total power consumption $P_{\rm tot}$ is given by \eqref{eq:Ptot}.

In this paper, our objective is to maximize the long-term EE across multiple possible UE locations. We identify sparse array architectures and antenna selection from within such arrays as promising approaches for enhancing EE compared to denser array configurations.

\section{Why Sparse Arrays Are Advantageous for Green 6G MIMO}

In this section, we show that \emph{sparse arrays}, with inter-antenna spacing much larger than the wavelength, offer strong potential for improving EE. By preserving the same aperture while using fewer antennas, they reduce the number of active RF chains and overall power consumption. Since the aperture—and thus the Fraunhofer distance—remains unchanged, the favorable near-field characteristics for multi-stream communication are largely maintained.

To this end, we adopt the following simulation parameters. The carrier frequency is set to $15$\,GHz, and the communication bandwidth is $B=100$\,MHz. The noise variance is computed as $N_0BF$, where $N_0=-204$\,dBW/Hz denotes the thermal noise spectral density, and $F=10$\,dB is the receiver noise figure. The geometry assumes a height difference of $10$\,m between the UE and the ELAA, a UE-to-ELAA distance of $50$\,m, and an azimuth angle of $\pi/4$ with respect to the array broadside direction. The path loss is modeled using an exponent of $2.5$ and a reference path loss of $-60$\,dB at a distance of $1$\,m. The UE is equipped with $M=4$ antennas.

The hardware-related parameters are set as follows: $\kappa=0.4$, $\mu=100$\,mW, $D_0=20$\,mW, $\upsilon=10^{-10}$\,J/sample, and $\eta=10^{-11}$\,J/bit. The power amplifier compression parameter $\rho$ is varied from $-0.3$ to $0$, where $\rho=0$ corresponds to the ideal distortionless case. All UPAs share the same physical area, while the number of antennas and their spacing are adjusted accordingly. For each configuration, the total input power to the power amplifiers is swept in $2$\,dB steps between $-30$\,dBW and $-10$\,dBW. The classical water-filling (WF) power allocation is employed, and the precoders are chosen as the right singular vectors of the MIMO channel. The operating point (total input power and the resulting WF power allocation) that yields the highest EE is recorded.

\begin{figure}[t]  
    \centering
    \includegraphics[width=0.4\textwidth]{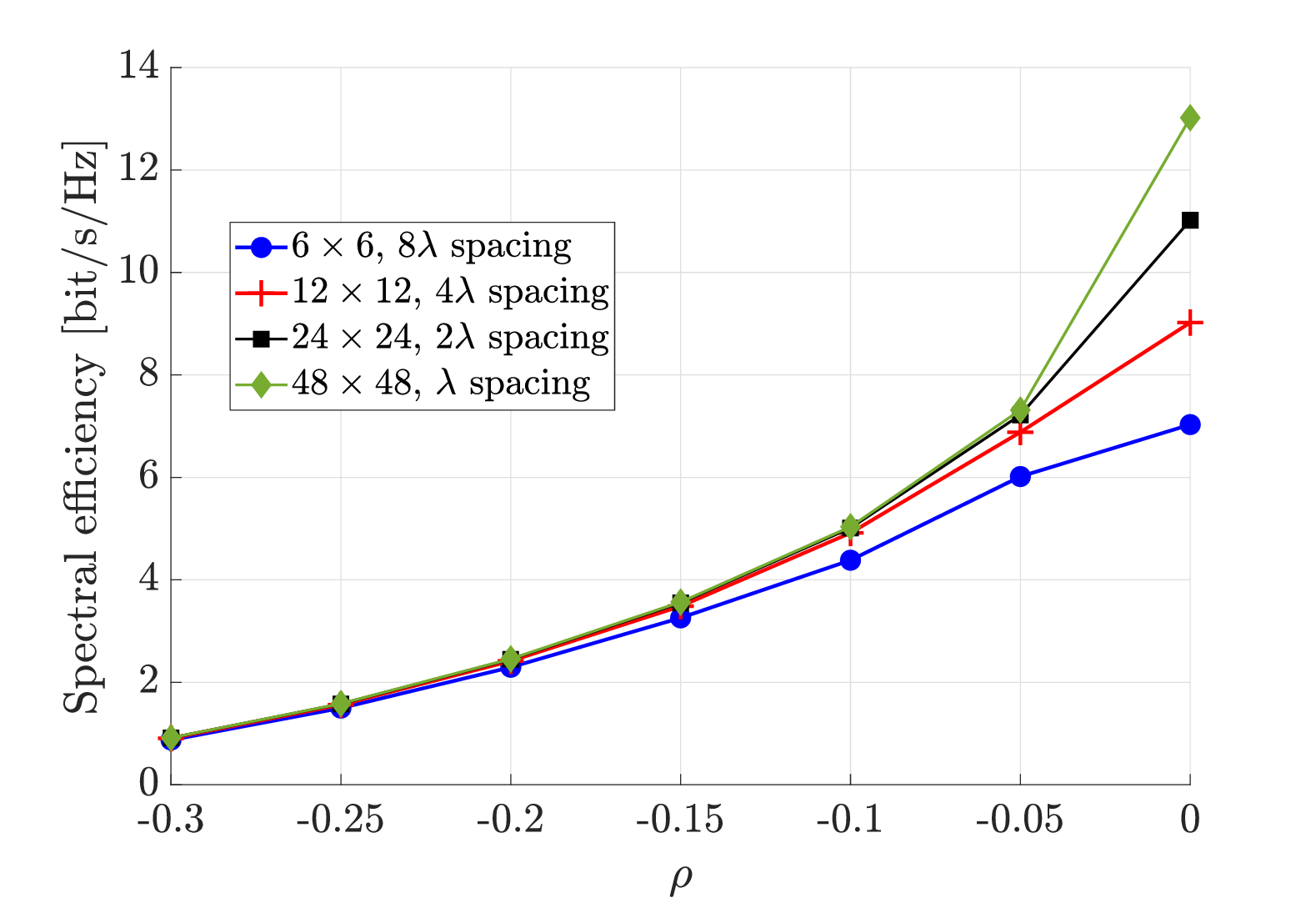}
        \vspace{-3mm}
    \caption{SE versus power amplifier compression parameter $\rho$ for different ELAAs.}
    \vspace{-5.5mm}
    \label{fig:SE-sparse}
\end{figure}
\begin{figure}[t]  
    \centering
        \includegraphics[width=0.4\textwidth, 
        ]{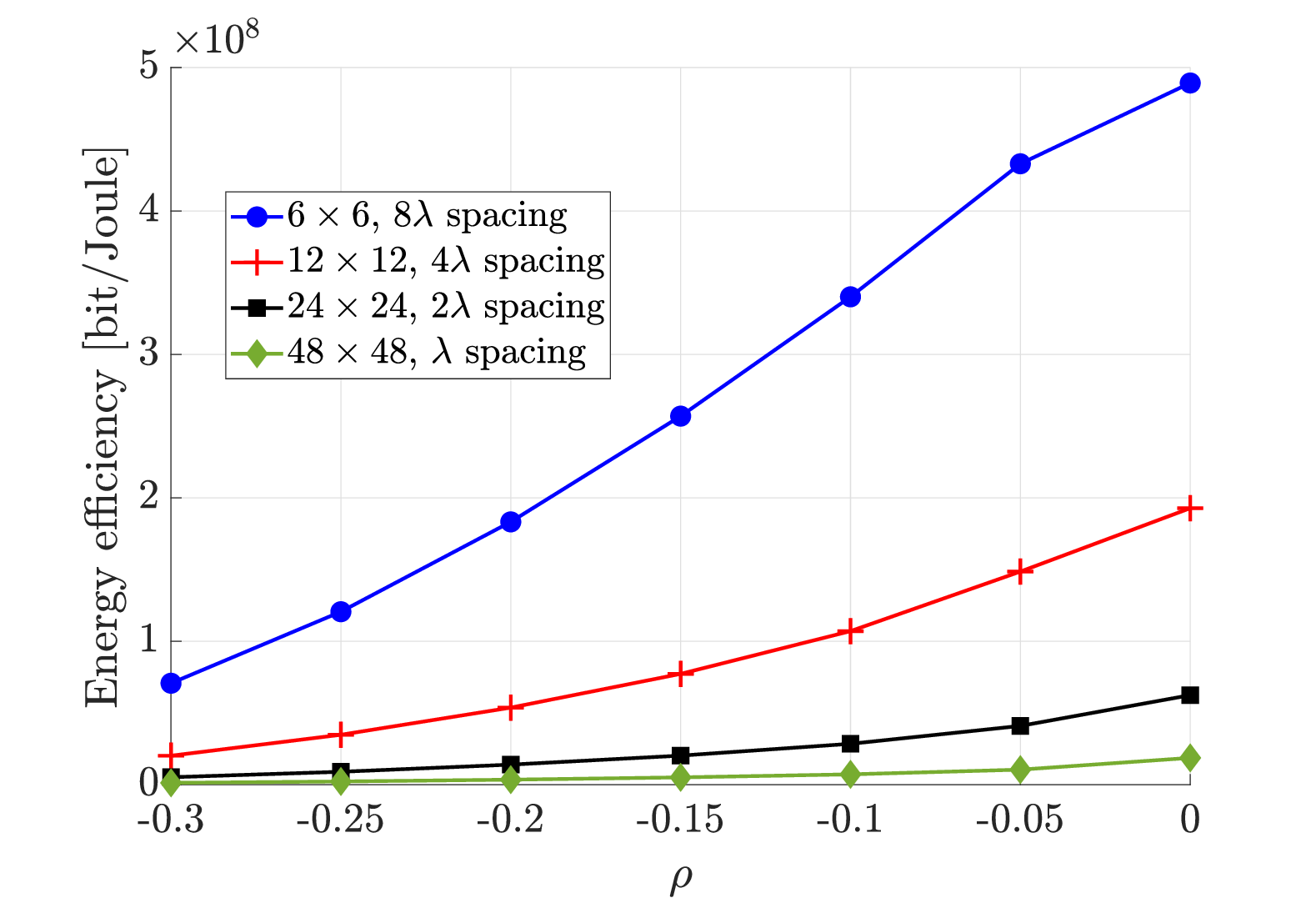}
            \vspace{-3mm}
        \caption{EE versus power amplifier compression parameter $\rho$ for different ELAAs. }
        \label{fig:EE-sparse}
            \vspace{-6mm}
 \end{figure}

Fig.~\ref{fig:SE-sparse} shows the SE variation with the compression parameter $\rho$ for different ELAA configurations. As expected, the densest array attains the highest SE due to its larger number of antennas. However, the SE gaps shrink under strong distortion. Furthermore, as shown in Fig.~\ref{fig:EE-sparse}, the sparsest array consistently achieves the highest EE across all $\rho$ values, outperforming its denser counterpart. Specifically, under $\rho = -0.05$, a $6 \times 6$ sparse ELAA attains nearly the same performance—within only 1 bit/s/Hz—compared to a much denser $48 \times 48$ array, while providing almost a fivefold improvement in EE. Therefore, in the next section, we focus on a sparse $6 \times 6$ ELAA with an inter-antenna spacing of $8\lambda$ as the reference configuration, and investigate antenna selection to further improve EE.

\section{Search Space Reduction via Sparse Modular Array Antenna Selection}

In the previous section, we focus on the numerical analysis for a single UE. However, the optimal EE operating point depends on the UE’s location and the corresponding channel characteristics. In some cases, activating more antennas within the available sparse array improves EE, while in other cases, deactivating a subset of antennas  yields a higher EE. A brute-force search over all possible antenna activation combinations would be computationally prohibitive. To address this challenge, we draw inspiration from the concept of \emph{modular arrays} in the literature. Specifically, we partition the $6\times 6$ sparse array into four modular sub-arrays, each forming a $3\times 3$ antenna block located at the edges of the overall aperture. For a given configuration, specified by a pair of parameters that determine the active sub-array dimensions (e.g., $(2,2)$ corresponds to a $2\times 2$ active antenna block), our goal is to select which antennas to activate. Within each modular sub-array, the active antennas are restricted to configurations that conform to either a UPA or a uniform linear array (ULA).

\begin{figure}[t]  
    \centering
        \includegraphics[width=0.4\textwidth]{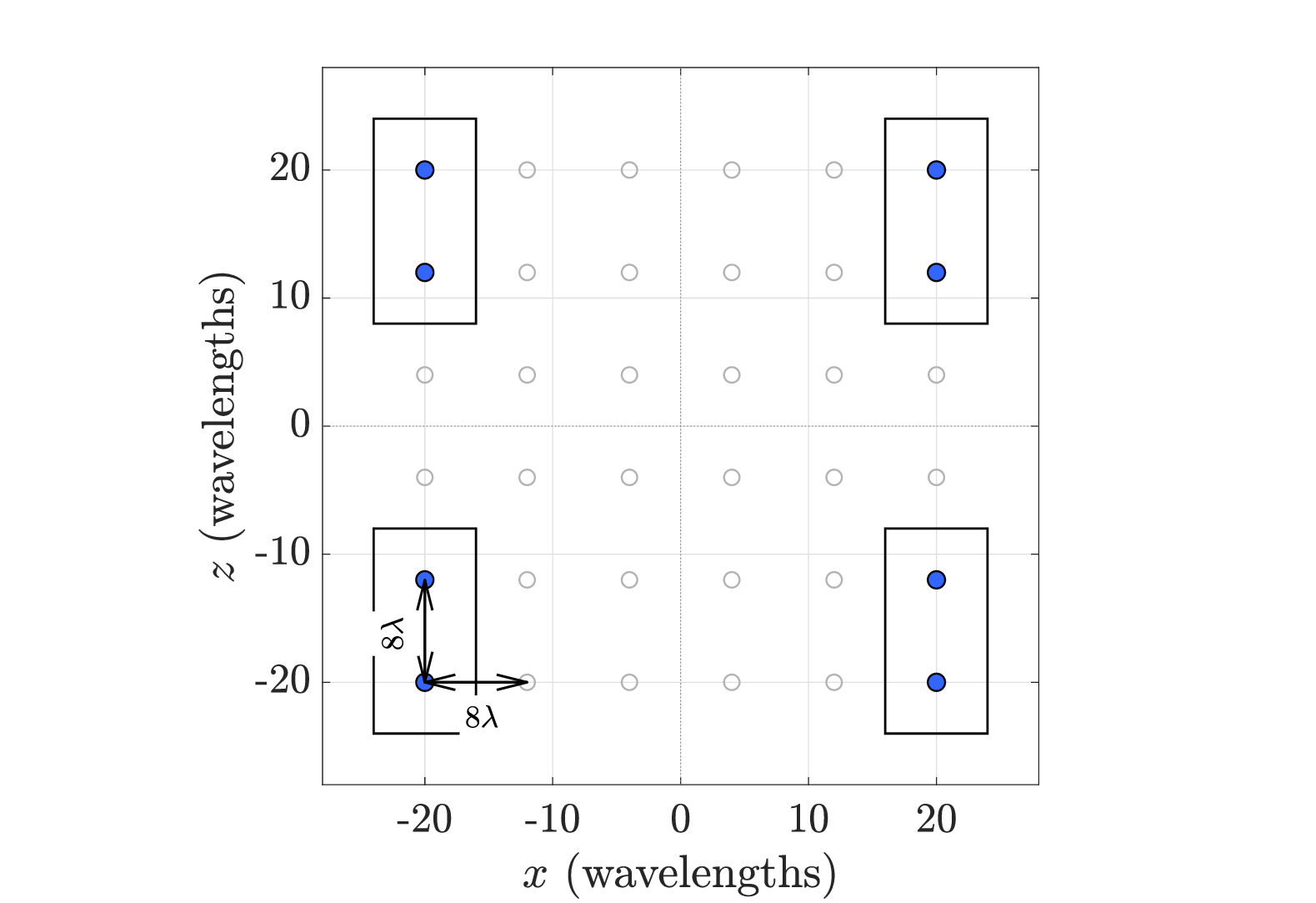}
        \caption{Example of the modular array structure with $C_x=1$ and $C_y=2$, where a $1\times2$ vertical ULA is activated in each of the four corner sub-arrays.}
        \label{fig:selection}
\end{figure}

For a  $6\times 6$ sparse array, the first configuration parameter $C_x \in \{1,2,3\}$ specifies the number of horizontally active antennas in each modular array, while $C_y \in \{1,2,3\}$ determines the number of vertically active antennas. For instance, in Fig.~\ref{fig:selection}, $C_x=1$ and $C_y=2$, meaning that a $1\times2$ vertical ULA is activated within each modular sub-array. In total, there are $K=2\times4=8$ active antennas. Notice that the antennas located at the outer edges are always active to maintain the overall aperture. The main motivations behind this modular activation rule are:
\begin{itemize}
    \item To preserve the overall array aperture, ensuring minimal performance loss while benefiting from reduced hardware usage.
    \item To minimize the optimization parameters, avoiding high computational complexity. Note that there are only two discrete parameters in this approach, $C_x$ and $C_y$, rather than exhaustive combinatorial searches. 
\end{itemize}

\section{DNN-Based Energy-Efficient Power Allocation and Antenna Selection}

We begin by noting that, even for a single UE location and without considering EE, maximizing the SE is already a non-trivial task. This is due to the complicated structure of the SE expression in \eqref{eq:SE} under power-amplifier non-linearities, which makes the classical WF solution suboptimal. When EE maximization is included, the problem becomes even more difficult because antenna selection introduces a combinatorial search. 
To address this, we propose a DNN-based method that learns the non-linear mapping from channel state information to the EE-optimal operating point, jointly estimating the power allocation, transmit power scaling, and modular sub-array activation through softmax-based probabilities derived from geometric and singular-value features.

\subsection{Input Feature Construction}
Each training sample corresponds to a UE geometry parameterized by range and azimuth $(r,\varphi)$ and its associated near-field channel matrix $\mathbf{H}\!\in\!\mathbb{C}^{M\times \overline{K}}$, where for our array $\overline{K}=6\cdot 6=36$.  
For every geometry, we compute the \emph{compact} singular value decomposition (SVD)
\begin{equation}
    \mathbf{H}=\mathbf{U}\boldsymbol{\Sigma}\mathbf{V}^{\Htran},
\end{equation}
with $\mathbf{U}\!\in\!\mathbb{C}^{M\times M}$, $\boldsymbol{\Sigma}\!\in\!\mathbb{R}^{M\times M}$, and $\mathbf{V}\!\in\!\mathbb{C}^{\overline{K}\times M}$, where $M\leq\overline{K}$. We denote the singular values by $\{s_i\}_{i=1}^M$. The DNN input feature vector is then
\begin{equation}
    \mathbf{f}
    =
    \begin{bmatrix}
        \ln(1+s_1)\\ \vdots\\ \ln(1+s_{M})\\[1mm]
        r\\ \varphi\\[1mm]
        \Re\!\big(\mathrm{vec}(\mathbf{V})\big)\\
        \Im\!\big(\mathrm{vec}(\mathbf{V})\big)
    \end{bmatrix}\!,
\end{equation}
where the $\log(1+s_i)$ transformation compresses the dynamic range of the singular values and was empirically observed to improve training performance. The motivation for including only the right singular vectors is that the effect of the left singular matrix can be cancelled in the SE expression \eqref{eq:SE} by multiplying the terms inside the determinant and inside the inverse by $\vect{U}^{\Htran}$ from the left and by $\vect{U}$ from the right.

\subsection{Dataset Generation and Spatial Sampling}

The UE locations $(r, \varphi)$ are first uniformly generated within a circular region on the $x$–$y$ plane, and only those falling inside the first quadrant are retained in the dataset. This restriction is justified by the geometric symmetry of the channel with respect to the second quadrant in front of the array, where only the azimuth angle changes. Therefore, the azimuth angle is limited to the interval $\varphi \in [0, \pi/2]$ to avoid redundant channel realizations.
Since the height difference between the ELAA and the UEs is fixed, the elevation angle is uniquely determined for each $(r,\varphi)$ pair.

For each valid location, the corresponding near-field channel matrix is constructed as
\begin{equation}
[\mathbf{H}]_{m,k} =
\frac{\sqrt{\beta_0}}{d_{mk}^{\xi/2}}
e^{-\imagunit 2\pi d_{mk}/\lambda},
\end{equation}
where $d_{mk}$ denotes the distance between the $k$th transmit and the $m$th receive antenna, $\xi$ is the path-loss exponent, and $\beta_0$ is the reference path loss at a distance of one meter. The wavelength is denoted by $\lambda$.

During training, each sample contributes to the overall loss with a weight proportional to $r_j^{-\alpha}$, i.e.,
\begin{equation}
\mathcal{L}(\boldsymbol{\theta}) =
-\frac{1}{S}\sum_{j=1}^{S} 
\frac{\mathrm{EE}_j(\boldsymbol{\theta})}{r_j^\alpha},
\end{equation}
where $r_j$ denotes the distance of the $j$th UE in the mini-batch, $S$ is the mini-batch size, and $\boldsymbol{\theta}$ represents the learnable parameters.
This distance-based weighting compensates for the geometric imbalance inherent in circular sampling—since the area element increases with $r$, fewer samples naturally appear near the center.
While a strict $1/r_j$ weighting would overcompensate this imbalance, a sub-linear exponent $r_j^{\alpha}$ with $0<\alpha<1$ provides a smoother trade-off.
This choice gently emphasizes distant UEs, whose EE is typically lower, without excessively diminishing the influence of nearby ones.
Consequently, the DNN learns a more balanced and representative power-allocation policy across the entire coverage area.

\subsection{Network Architecture and Activation Functions}

The DNN is designed as a feed-forward, fully connected architecture consisting of several hidden layers.
Each hidden layer employs the \emph{Sigmoid Linear Unit} (SiLU) activation function, defined as
\begin{equation}
\mathrm{SiLU}(x) = x\sigma(x), \qquad
\sigma(x) = \frac{1}{1+e^{-x}},
\end{equation}
where $\sigma(x)$ denotes the logistic sigmoid function.
Unlike ReLU, the SiLU activation is smooth and differentiable over the entire real line, and we empirically observed that it leads to better training performance.

The final layer branches into four output heads.
The first head produces the power allocation logits, followed by a softmax activation with $M$ outputs corresponding to the parallel spatial subchannels of the MIMO channel.
The second head applies a sigmoid activation to generate a scalar power-scaling coefficient that regulates the total transmit power.
These two outputs are combined to obtain the final power coefficients as
\begin{equation}
\vect{p} = P_{\rm max}\sigma(u)\mathrm{softmax}(\vect{p}_{\mathrm{logits}}),
\label{eq:p_scaling}
\end{equation}
where $P_{\rm max}$ is the maximum total input power to the power amplifiers, $\vect{p}_{\mathrm{logits}}$ represents the unnormalized DNN outputs, and $\sigma(u)\in(0,1)$ controls the power reduction factor for achieving distortion-aware and energy-efficient operation.
Hence, the DNN jointly learns both the relative power allocation among subchannels (via the softmax) and the total transmit power scaling (via the sigmoid), ensuring that the resulting coefficients are non-negative and bounded by $P_{\rm max}$.

The remaining two output heads correspond to the modular sub-array activation probabilities along the horizontal and vertical dimensions, denoted by $\mathbf{P}_x$ and $\mathbf{P}_y$, respectively.
Each is obtained through a softmax layer with three outputs.
The indices of the maximum entries of $\mathbf{P}_x$ and $\mathbf{P}_y$ determine the horizontal and vertical activation parameters $(C_x, C_y)$, representing the number of active antennas in each modular sub-array.

The transmit covariance matrix is expressed as
\begin{equation}
\vect{Q} = \vect{V}\vect{P}\vect{V}^{\Htran},
\end{equation}
where $\vect{V}\in\mathbb{C}^{K\times M}$ contains the right singular vectors of the channel matrix $\vect{H}$ (with $K$ denoting the number of active transmit antennas, noted here with a slight abuse of notation), and $\vect{P} = \mathrm{diag}(p_1, p_2, \ldots, p_M)$ is the diagonal power-allocation matrix whose entries are defined according to~\eqref{eq:p_scaling}. Note that the channel matrix $\vect{H}$ has dimensions $M\times K$, where $K$ denotes the number of active transmit antennas.
In contrast, the input feature to the DNN is constructed using the full channel of size $M\times \overline{K}$, where $\overline{K}$ represents the total number of available antenna elements.

\section{Numerical Results}

In this section, the SE and EE performances of the proposed DNN-based approach are quantitatively evaluated against several benchmarks. The model, implemented in \texttt{PyTorch} and trained on an NVIDIA GPU, adopts a seven-layer feed-forward architecture with widths $(64,128,128,128,128,64,64)$, \textit{SiLU} activations, and a dropout rate of $0.1$. Training uses the AdamW optimizer ($\text{lr}=10^{-3}$, weight decay $=2\times10^{-4}$) with cosine-annealing scheduling (min $\text{lr}=10^{-5}$) over $10$ epochs. The dataset includes $N_{\mathrm{train}}=5000$ and $N_{\mathrm{val}}=2000$ user geometries. Each mini-batch contains $64$ samples, and the loss is the negative weighted EE, scaled by $1/r_j^{0.6}$ to balance near and far UEs.

The UE distances and azimuth angles are uniformly selected from the ranges $[10,200]$\,m and $[0,\pi/2]$, respectively. 
The hardware parameters are the same as described earlier, and the maximum total transmit power is set to $P_{\rm max}=250$\,mW. 
The number of UE antennas is $M=4$. 
For comparison, we consider four additional benchmark schemes:
\begin{itemize}
    \item \textbf{WF (with learned power \& antennas):} 
This scheme applies classical WF power allocation using a total transmit power level $P_{\rm max}\sigma(u)$, where $\sigma(u)$ is the power-scaling factor learned by the DNN. The active antenna subset is also selected based on the DNN-predicted modular configuration.
\item \textbf{WF-All (with learned power):} 
In this case, WF is applied over the entire $6\times6$ array (i.e., all $36$ antennas are active), while the total power level $P_{\rm max}\sigma(u)$ is taken from the DNN output.
\item \textbf{WF (with learned antennas \& full power):} 
Here, WF-based power allocation is performed at the full power level $P_{\rm max}$, whereas the active antenna configuration is determined by the DNN.
\item \textbf{WF-All (with full power):} 
This reference represents the conventional WF solution with all $36$ antennas active and the full transmit power $P_{\rm max}$ applied.

\end{itemize}

\begin{figure}[t]  
    \centering
        \includegraphics[width=0.4\textwidth, 
        ]{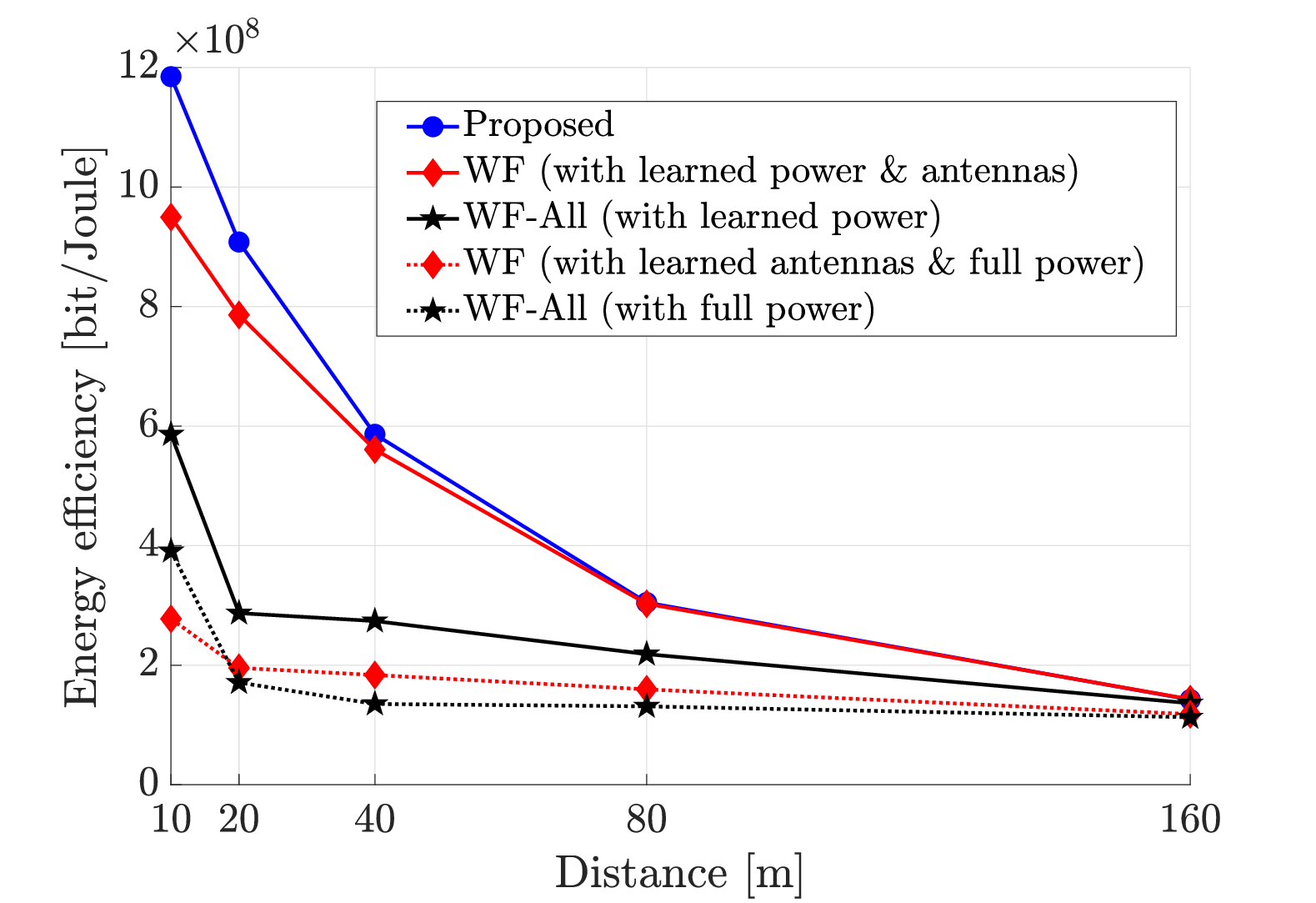} 
            \vspace{-2mm}
        \caption{EE comparison between the proposed DNN-based method and benchmark schemes.  }
        \label{fig:EE-comparison}
            \vspace{-2mm}
 \end{figure}

In Fig.~\ref{fig:EE-comparison}, the proposed is shown to achieve the highest EE across all user distances by jointly learning distortion-aware power allocation, transmit power scaling, and antenna activation. The ``WF-All (full power)'' baseline suffers from excessive power usage due to full antenna activation, while the ``WF (learned power \& antennas)'' scheme closely tracks the DNN, demonstrating good generalization of the learned parameters. At short distances, power allocation effects become more pronounced, resulting in a slight performance gap.

\begin{figure}[t]  
    \centering
        \includegraphics[width=0.4\textwidth, 
        ]{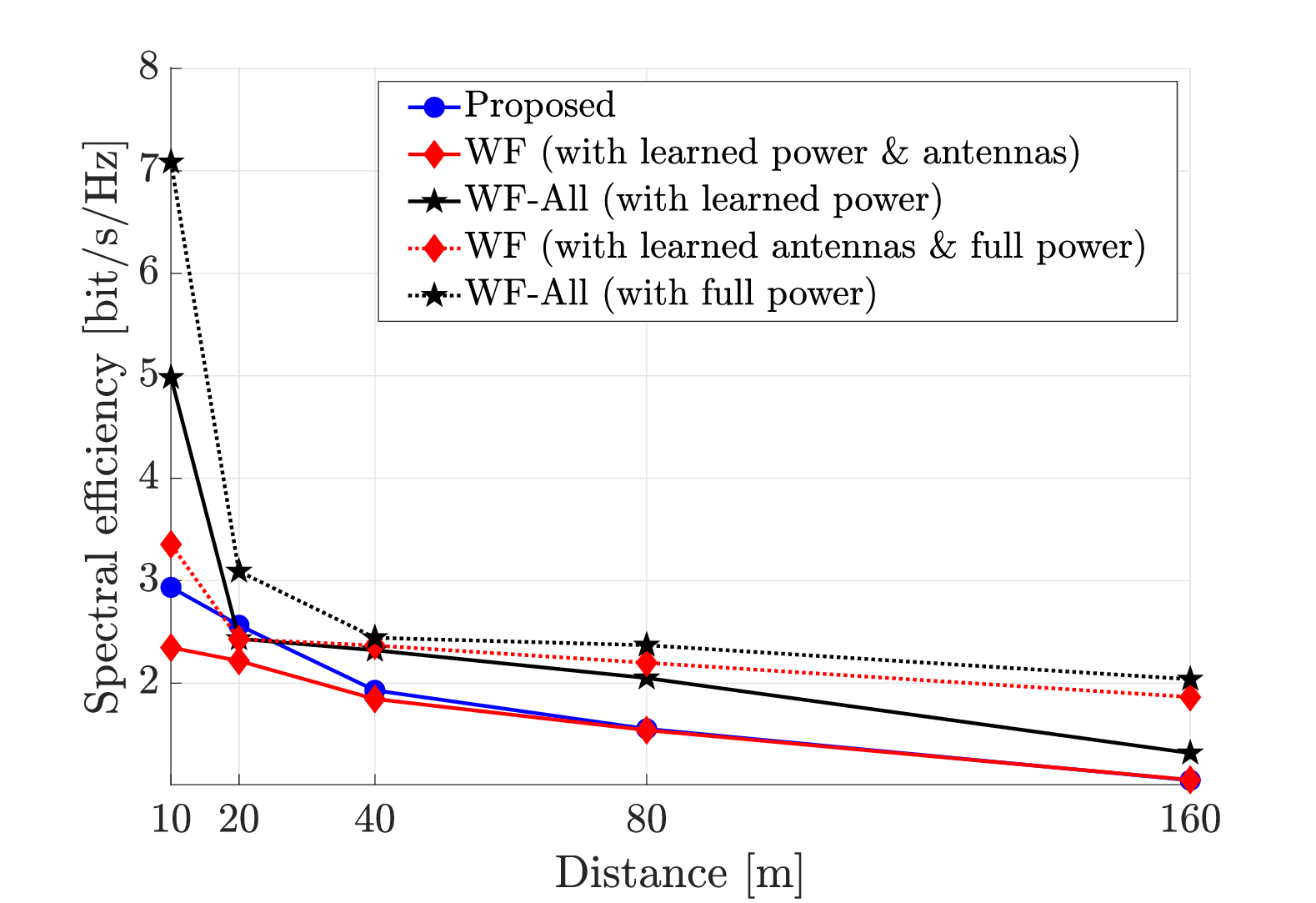} 
            \vspace{-2mm}
        \caption{SE comparison between the proposed DNN-based method and benchmark schemes. }
        \label{fig:SE-comparison}
            \vspace{-2mm}
 \end{figure}

In Fig.~\ref{fig:SE-comparison}, we observe that the DNN-based approach attains nearly identical SE except at very close UE distances, while providing much higher EE, confirming its energy-aware behavior. The WF (learned antennas \& full power)'' and ``WF-All (learned power)'' schemes achieve intermediate performance, demonstrating that learning only antenna activation or power scaling provides partial gains, whereas the proposed joint learning approach offers the best SE–EE trade-off.

\vspace{-2mm}
\section{Conclusion}
\vspace{-2mm}
This paper presented an unsupervised DNN-based framework for energy-efficient power allocation and modular antenna activation in MIMO systems with non-linear power amplifiers.
By exploiting Bussgang-based distortion modeling and sparse array structures, the proposed approach learns a joint policy for transmit power scaling, spatial power allocation, and modular sub-array activation.
The results highlight that the DNN generalizes well across different UE geometries and effectively captures distortion-induced non-linearities, achieving superior EE compared to all benchmark schemes.
Furthermore, the study confirms the advantage of sparse modular arrays, which allow dynamic activation of sub-arrays to sustain high EE while maintaining near-optimal SE.
Unlike the conventional WF with full-array activation, which maximizes SE at the expense of power consumption, the proposed method adaptively selects antenna subsets and power levels to achieve the best SE–EE trade-off.
These findings establish sparse-array-based, distortion-aware learning as a promising direction for future energy-efficient 6G MIMO transceivers.

\bibliographystyle{IEEEtran}
\bibliography{IEEEabrv,refs}

\end{document}